\documentstyle[floats,aps,graphicx,prl]{revtex}

\begin{document}

\setlength{\unitlength}{1cm}

\draft
\twocolumn[\hsize\textwidth\columnwidth\hsize\csname @twocolumnfalse\endcsname

\title{Dispersion and Symmetry of Bound States in the Shastry-Sutherland Model}

\author{Christian Knetter, Alexander B\"uhler, Erwin M\"uller-Hartmann, 
and G\"otz S. Uhrig}

\address{Institut f\"ur Theoretische Physik, Universit\"at zu
  K\"oln, D-50937 K\"oln, Germany}

\date{\today}

\maketitle

\begin{abstract}
Bound states made from two triplet excitations
on the Shastry-Sutherland (ShaSu) lattice
are investigated. Based on the perturbative
unitary transformation by flow equations quantitative properties like
dispersions and qualitative properties like symmetries are determined.
The high order results (up to $(J_2/J_1)^{14}$) permit to fix the
parameters of SrCu$_2$(BO$_3$)$_2$ precisely: $J_1=6.16(10)$meV, 
$x:=J_2/J_1=0.603(3)$, $J_\perp=1.3(2)$meV. 
At the border of the magnetic Brillouin zone (MBZ) a general double degeneracy
is derived. An unexpected instability in the triplet channel at $x=0.63$
indicates a first order transition towards a triplet condensate,
related to classical helical order.
\end{abstract}
\pacs{PACS numbers: 75.40.Gb, 75.30.Kz, 75.50.Ee, 75.10.Jm}


\centerline{\it Dedicated to Prof. F. Wegner on occasion of his 60$^{\rm th}$ birthday.}
\medskip
 ]

Quantum
antiferromagnets are at the center of research
not only because of the high $T_c$ superconductors . Of particular
interest are systems which do not have an ordered, N\'eel-type
ground state. Their ground state is a spin liquid without long
range spin order.  Spin liquids  are favored by low spin ($S=\frac{1}{2}$
mostly), low coordination number ($Z\in\{2,3,4\}\Rightarrow D\in\{1,2\}$),
 and strong geometric frustration.

Dimer solids are  transparent cases of spin liquids.
In $D=1$, the generic example is the Majumdar-Ghosh model
\cite{majum69a} of which Shastry and Sutherland
found a $D=2$ generalization (ShaSu model) \cite{shast81b}. In both cases
frustration is essential. Each spin is coupled to pairs of
spins (dimers). If these pairs form singlets
the couplings between dimers is without effect and the
singlet-on-dimers product state is always an eigen state and for
certain parameters the ground state
\cite{broek80,shast81a,shast81b,mulle00a}. The systems are gapped.
The elementary excited states are 
 dressed $S=\frac{1}{2}$ ($D=1$) \cite{shast81a} or $S=1$
($D=2$) entities. They interact strongly and form bound and antibound 
states in various spin channels.
\begin{figure}[htbp]
  \begin{center}
    \begin{picture}(8,4.2)
      \put(0,0.2){
        \includegraphics[width=7.5cm]{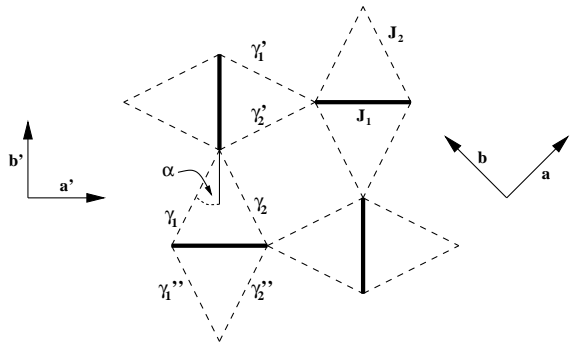}}
      \end{picture}
    \caption{Unit cell of the Shastry-Sutherland (ShaSu) lattice as realized
in SrCu$_2$(BO$_3$)$_2$, ({\bf a},{\bf b}) are unit vectors; ({\bf a'},{\bf b'})
and the coefficients $\gamma$ are used to analyse Raman scattering.}
    \label{fig:raman}
  \end{center}
\end{figure}
Due to its recent realization in
SrCu$_2$(BO$_3$)$_2$ \cite{kagey99,miyah99} the
 ShaSu model (Fig.~\ref{fig:raman}) is presently attracting
enormous interest. The Hamiltonian reads
\begin{equation}
\label{hamilton}
H(J_1,J_2)= J_1 \sum_{\langle {\bf i},{\bf j}\rangle {\rm dimer}} {\bf
S}_{{\bf i}}{\bf S}_{{\bf j}} + J_2 \sum_{\langle {\bf i},{\bf j}
\rangle {\rm square}}{\bf S}_{{\bf i}}{\bf S}_{{\bf j}} \ .
\end{equation}
In this Letter we start from the dimer phase \cite{mulle00a}. We focus
 on bound states formed from pairs of the
elementary triplets and their symmetries,
degeneracies, and dispersion. The perturbative
unitary transformation \cite{knett00a} based on flow equations \cite{wegne94}
enables us to link smoothly and uniquely $H(J_1,J_2)$ at
$x:=J_2/J_1 \neq 0$ to an effective  $H_{\rm eff}$ conserving the
number of  triplets on dimers $[H_{\rm eff},H(J_1,0)]=0$. This permits
a clear distinction between the ground state (without triplets),
 the 1-triplet sector, the 2-triplet sector etc..

In terms of $H_{\rm eff}$ the dynamics of one triplet is hopping
 $t_{h;{\bf i}}$ ($t_{v;{\bf i}}$) starting from 
a horizontal (vertical) dimer 
by $i_x$ dimers  right and $i_y$ dimers up. Nothing else is possible
due to triplet number conservation. The elements $t$
are computed in order 15 \cite{mulle00a,weiho99a,knett00c}.

The dynamics of two triplets  at large distances is governed by  1-triplet
hopping. At smaller distances a  2-particle
interaction occurs additionally given by $W_{h;{\bf d};{\bf i},{\bf d'}}$ 
($W_{v;{\bf d};{\bf i},{\bf d'}}$)
 starting with one triplet on a horizontal (vertical) dimer and another at
distance ${\bf d}$. The action of $H_{\rm eff}$ is to shift the triplets
to ${\bf i}$ and to ${\bf i}+{\bf d'}$. Nothing else is possible
due to  triplet number conservation. 
Since the total spin is conserved
($S\in\{0,1,2\}$) the distances are restricted to 
${\bf d}, {\bf d'} > {\bf 0}$, i.e.\ $d_x>0$ or $d_x=0 \wedge d_y>0$,
 because the exchange parity is fixed.

The action of $H_{\rm eff}$ yields the {\it combined} effect of  hopping
and  interaction denoted by $A_{{\bf d};{\bf i},{\bf d'}}$.
The true 2-triplet interaction is easily found by subtracting the
 1-triplet hopping \cite{mulle00a,weiho99a,knett00c}
\begin{mathletters}
\label{interpure}
\begin{eqnarray}
W_{{\bf d};{\bf 0},{\bf d'}} &=& A_{{\bf d};{\bf 0},{\bf d'}}
-t_{{\bf d'}-{\bf d}} -\delta_{{\bf d'},{\bf d}} t_{{\bf 0}} \\
W_{{\bf d};{\bf d}-{\bf d'},{\bf d'}} &=& 
A_{{\bf d};{\bf d}-{\bf d'},{\bf d'}}
-t_{{\bf d}-{\bf d'}} -\delta_{{\bf d'},{\bf d}} t_{{\bf 0}} \\
W_{{\bf d};-{\bf d'},{\bf d'}} &=& A_{{\bf d};-{\bf d'},{\bf d'}}
-t_{-{\bf d}-{\bf d'}} \\
W_{{\bf d};{\bf d},{\bf d'}} &=& A_{{\bf d};{\bf d},{\bf d'}}
-t_{{\bf d}+{\bf d'}} 
\end{eqnarray}
\end{mathletters}
(distinction $h/v$ omitted for clarity).
Otherwise $A$ and $W$ are identical.
The coefficients $W$ for $S\in\{0,1,2\}$ yield the complete 
2-particle dynamics. We compute $W$ up to $x^{12}$, the coefficients for 
 the lowest-lying  states even  up  to $x^{14}$.

 During the  virtual
processes \cite{note1}  the triplet number is  changed.
Due to the  frustration of the ShaSu lattice
 $H$ in (\ref{hamilton})
changes the number of triplets on the dimers at most by one
\cite{miyah99,weiho99a,knett00c}.
An excitation or a de-excitation
 on a horizontal (vertical) dimer is possible iff at least one of the vertical
 (horizontal) dimers on the left and right (above and below) are excited.
This restriction implies that one triplet hops only in $x^6$ (cf.~Figs.\
in \cite{miyah00a,knett00c}).

\begin{figure}[htbp]
  \begin{center}
    \begin{picture}(8,3.2)
      \put(0,2){
        \includegraphics[width=8cm]{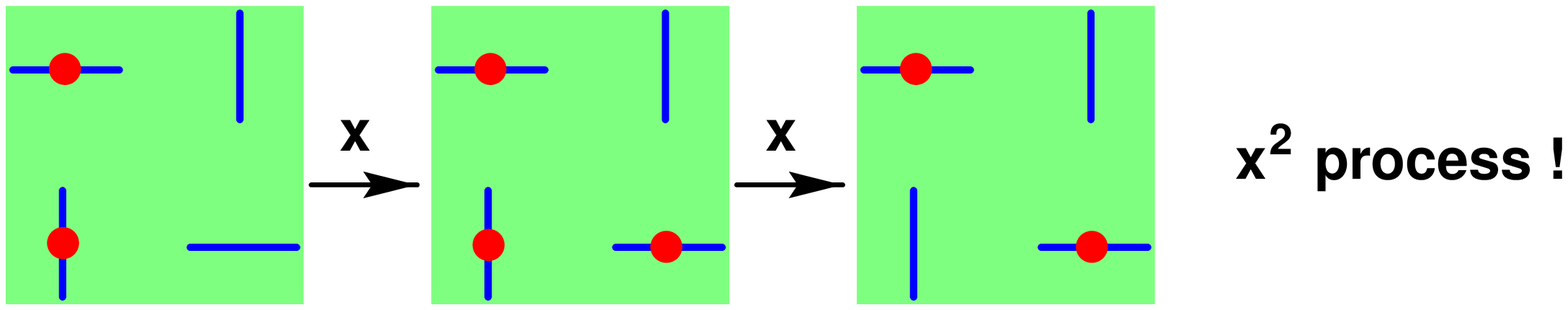}}
      \put(0,0.2){
        \includegraphics[width=8cm]{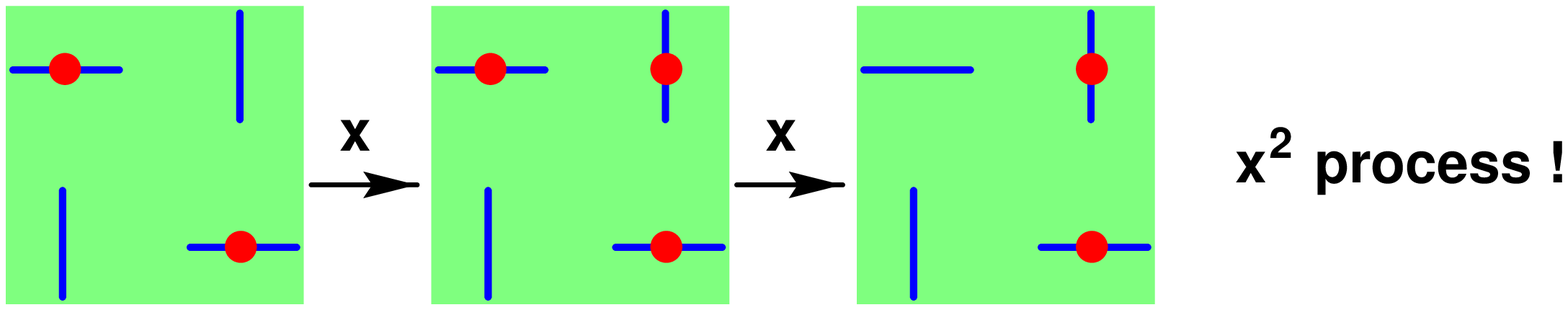}}
    \end{picture}
    \caption{Leading processes of correlated 2-triplet hopping. Dark dots are
      triplets, bars are dimers.}
    \label{fig:two-trip}
  \end{center}
\end{figure}

Motion of two triplets together is much less restricted 
(cf.~Fig.~\ref{fig:two-trip}). Matrix elements occur in $x^2$ as first 
observed for total spin $S=2$ \cite{momoi00}. But the dispersion of bound states
 starts only in  $x^3$
(contrary to $x^4$ claimed in Ref.~\cite{lemme00a}). 
Two adjacent triplets interact linearly in $x$
($-x$ for $S=0$, $-x/2$ for $S=1$, $x/2$ for $S=2$). The energy  of the
initial and final state in each row in
 Fig.~\ref{fig:two-trip} differ by ${\cal O}(x)$.
Hence both rows must be combined making it an  $(x^2)^2/x = x^3$ process eventually.
 This  applies
to the 8 (anti)bound states derived from two triplets on (next) nearest neighbor
 dimers. The dispersion of any other state sets in 
at higher order.

We use the following basis for the 2-triplet states
\begin{equation}
\label{basis}
|{\bf k},{\bf d},\sigma\rangle:=N^{-1/2}
\sum_{\bf r} 
e^{\left(i({\bf k}+\sigma{\bf Q})({\bf r}+{\bf d}/2)\right)}
 |{\bf r},{\bf r}+{\bf d}\rangle\ ,
\end{equation}
where ${\bf k}$ is the conserved total momentum in the magnetic
Brillouin zone (MBZ) applying due to the two sublattices;
$\sigma\in\{0,1\}$, ${\bf Q}:=(\pi,\pi)$, $N$ is the 
number of dimers, $|{\bf r},{\bf r}+{\bf d}\rangle $ denotes the state with triplets at
${\bf r}$ and at ${\bf r}+{\bf d}$. The distance ${\bf d}$ is restricted
${\bf d}>{\bf 0}$, i.e.\ $d_x>0$ or $d_x=0 \wedge d_y>0$.
 The matrix elements
of $H_{\rm eff}$ in the basis (\ref{basis}) are real due to
 translation and inversion symmetry.

Before  the quantitative analysis of the bound states 
a qualitative aspect, a general double degeneracy at the border of the MBZ, shall
be derived. To see this consider
 the combination of a shift by the dimer-dimer  
spacing along {\bf a'} (S), a reflection about {\bf a} (R), and the inversion
${\bf r} \to -{\bf r}$ (I) (cf. Fig.~\ref{fig:raman}). The combinations
SR and I are  symmetries of the Hamiltonian. For $k_x+k_y=\pi$
(part of the MBZ border)
definition (\ref{basis}) implies for the total combination SRI
the mapping
\begin{equation}
\label{vsi}
|{\bf k},{\bf d},\sigma\rangle \to 
e^{ik_x+i\pi(d_x+d_y+\sigma+PS)}
|{\bf k},-(\begin{array}{c}d_y\\
d_x\end{array}),1-\sigma\rangle \ ,
\end{equation}
where  $P\in\{0,1\}$ being unity iff $-(d_y,d_x) <{\bf 0}$
so that the triplets must be swapped to pass from $-(d_y,d_x)$
to $(d_y,d_x)$. It is crucial that SRI links
$|{\bf k},(d_x,d_y),0\rangle$ and 
$|{\bf k},-(d_y,d_x),1\rangle$ like a 2D rotation 
$\left(\begin{array}{cc} 0 &-1\\ 1 & 0\end{array}\right) $
up to a prefactor. Hence its eigen vectors are complex with
 linearly independent real and imaginary part and so are the 
simultaneous eigen vectors of SRI
and $H_{\rm eff}$. Because $H_{\rm eff}$  is real 
the real and the imaginary part constitute in fact 
linearly independent eigen vectors to the
same eigen value. The same double degeneracy  is concluded for the other parts of the
MBZ border by S and 90$^\circ$ rotation (D). It is also valid in the 1-triplet
sector \cite{knett00c}.

The double degeneracy at the MBZ border is interesting for analysing experiment, too.
Degeneracy reduces the large  number of
energetically close states  helping to resolve different bound states.

Since 1-triplet hopping is of higher order than  interaction
 an analytic expansion for the energies of 
the bound states is possible. At finite order in $x$  only
configurations contribute where the two triplets are
not too far away from each other.
 Of course, higher orders imply larger, but still finite distances.
 In particular, the energies of the four states which evolve from 
neighboring triplets can be computed very well since their
interaction is linear. 
Investigating the
matrix elements shows that it is sufficient to study the distances
${\bf d}\in\{(0,1),(1,0),(1,\pm1)\}$ 
for order 5. To $x^{14}$ only 
${\bf d}\in\{(1,\pm2),(2,\pm1),(0,2),(2,0),(2,\pm2)\}$
must be added. So, for given total momentum only a finite 
$8\times 8$ or $24\times 24$ matrix has to be analysed.
For illustration consider  the elements $A_{(0,1);{\bf i},(2,1)}$
(the Fourier transform of  ${\bf i}$ yields the momentum dependence.)
connecting $(0,1)$ and $(2,1)$ which is ${\cal O}(x^4)$.
 By second order
perturbation one sees that the resulting energy shift is $(x^4)^2/x=x^7$ only.

Furthermore,
the  elements connecting shorter distances to longer distances
and the elements among longer distances do not need to be known to 
very high order. Consider again the process 
 $(0,1) \leftrightarrow (2,1)$. In order $x^7$ the element 
$A_{(0,1);{\bf i},(2,1)}$ must be known only in $x^4$ and
$A_{(2,1);{\bf i},(2,1)}$ only in $x^1$; in order $x^9$ the element 
$A_{(0,1);{\bf i},(2,1)}$ 
must be known only up to  $x^6$ and
$A_{(2,1);{\bf i},(2,1)}$ only in $x^3$ and so on.

We have analysed the dispersions in $x^5$
of the four  states bound linearly in $x$ in the MBZ. 
Fukumoto's results are mostly confirmed \cite{fukum00,note3}.
At particular points of high point group symmetry 
($(0,0)$,$(0,\pi)$,$(\pi/2,\pi/2)$)
 the  Hamiltonian splits into six blocks corresponding
to different representations of the square  point group 4mm. At these points
the analysis up to $x^{14}$ is carried out \cite{note3}.
 The symmetries are classified according to the
irreducible representations (four 1D, one 2D) 
of the point group 4mm
$\Gamma_1 (1), \Gamma_2 (x^2-y^2), \Gamma_3 (xy), \Gamma_4 (xy(x^2-y^2)), 
\Gamma_5 (x, y)$ where simple polynomials are given in brackets to show
 the transformation behavior.

\begin{figure}[htbp]
  \begin{center}
    \includegraphics[width=\columnwidth]{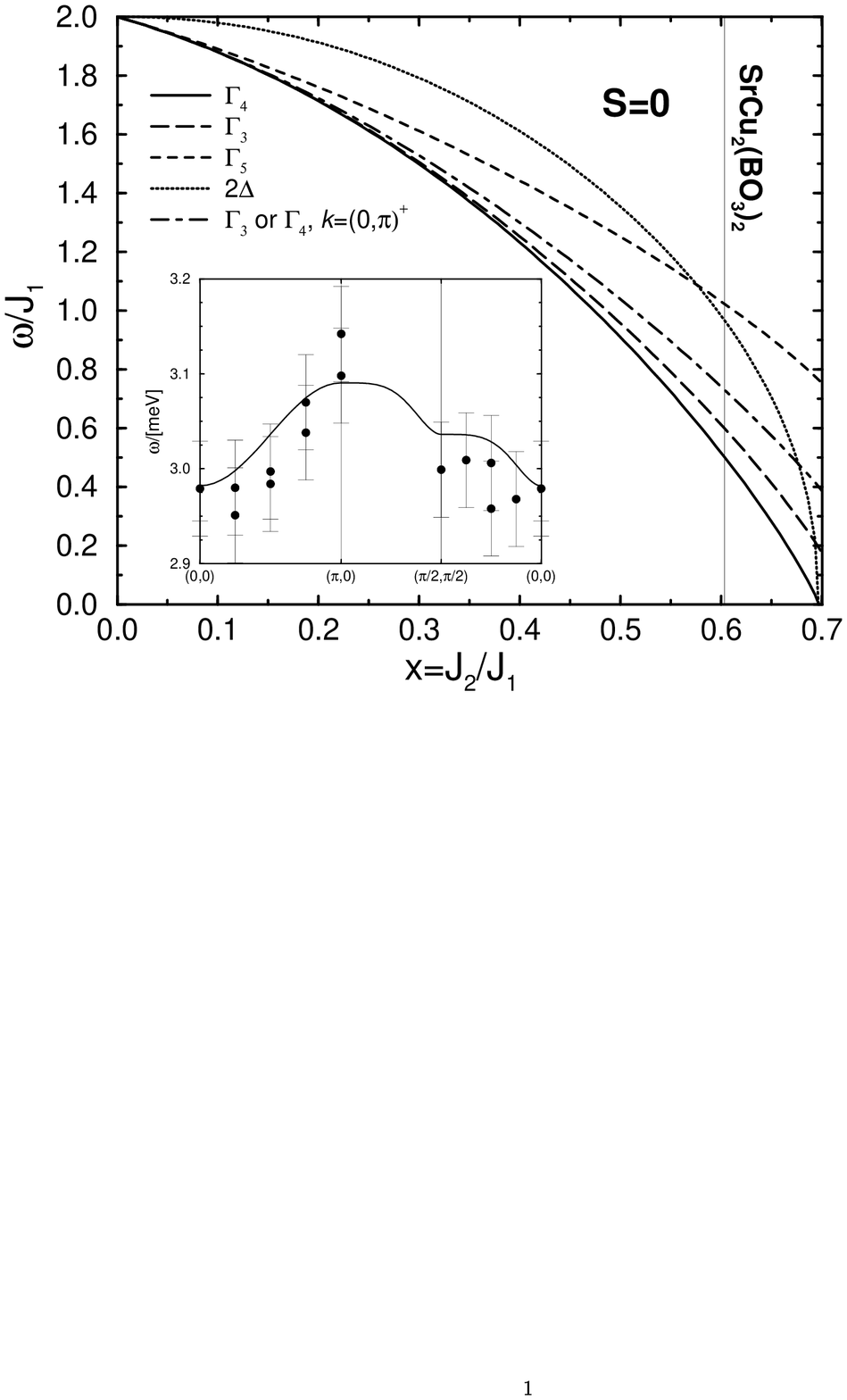}
    \caption{Energy of the lowest-lying $S=0$ states.
      Curves refer to ${\bf k}={\bf 0}$ except the
      dashed-dotted one. The dotted curve displays the continuum
      at $2\Delta$. Inset: 1-triplet dispersion. Theory
      at $x=0.603, J_1=6.16$meV; data from Ref.~\protect\cite{kagey00a},
      experimental errors at least as large as indicated.
      }
    \label{fig:s0}
  \end{center}
\end{figure}

\begin{figure}[htbp]
  \begin{center}
    \includegraphics[width=\columnwidth]{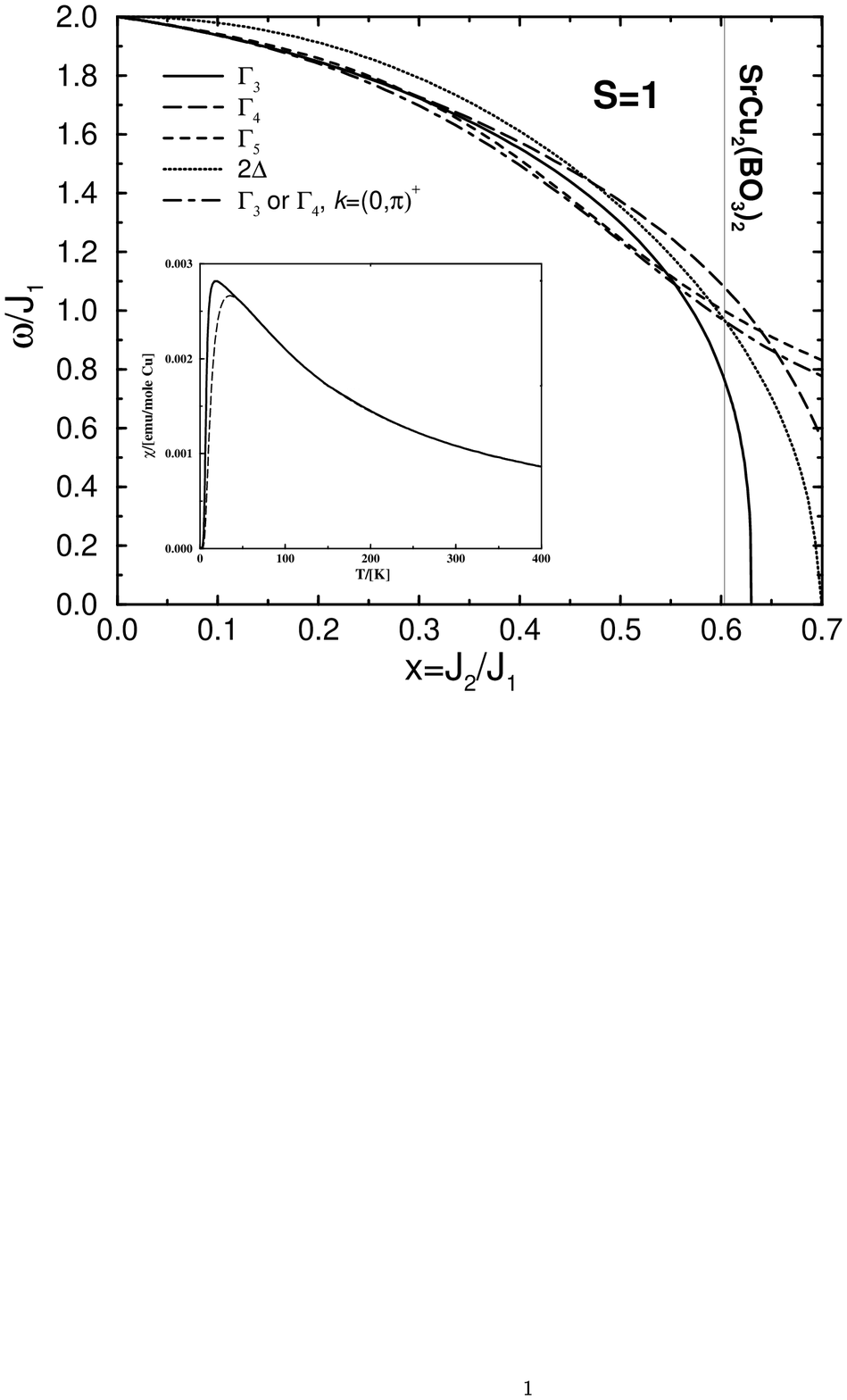}
    \caption{As in Fig.~\protect\ref{fig:s0} for  $S=1$.
      Inset: Magnetic susceptibility. Theory (dashed)
      with directional rms average $g=2.13$ \protect\cite{nojir99},
      $x,J_1$ as in Fig.~\protect\ref{fig:s0};
      experiment (solid) on powder \protect\cite{kagey99}.
      }
    \label{fig:s1}
  \end{center}
\end{figure}

The extrapolated energies are depicted in Figs.~\ref{fig:s0} ($S=0$)
 and \ref{fig:s1} ($S=1$) as functions of $x$.
For those energies which stay separated from the 2-particle
continuum  Dlog-Pad\'e approximants are used successfully  
\cite{domb89}. 
The results are stable under changes of the polynomial degrees.
 The energies close to the continuum (here simply
twice the gap $\Delta$ between the ground state and 
the elementary triplet at ${\bf k}={\bf 0}$) 
are given with less reliability
by the truncated series or by a non-defective Dlog-Pad\'e approximant.

In Figs.~\ref{fig:s0}, \ref{fig:s1} the modes are sorted in energetically
ascending order for small values of $x$: solid, long dashed, and short dashed 
curves. The $\Gamma_5$ modes are naturally degenerate. 
The double degeneracy for ${\bf k}=(0,\pi)$ does {\it not}
result from the point group but originates from the complex conjugation
as explained above. The dashed-dotted curve at $(0,\pi)$ has to
be compared to the solid and the long-dashed curve to assess the 
dispersion of these two modes from ${\bf 0}$ to $(0,\pi)$.
While for $S=0$ this dispersion always has the expected
behavior with $\omega({\bf 0}) < \omega((0,\pi))$ the energies for $S=1$
are reversed  for small values of $x$ (cf. \cite{fukum00}). 
Only above $x\approx 0.55$ the relation 
$\omega({\bf 0}) < \omega((0,\pi))$ holds for $S=1$.

We do not agree with Ref.~\cite{fukum00} that the
two lowest states are of  $s$-wave type since this 
would imply that they are $\Gamma_1$. Instead the $S=0,1$ states are 
odd under  reflection about {\bf a'} (R1) or about {\bf b'} (R2)
(cf.~Fig.~\ref{fig:raman}). 
For $S=0$, the lowest
state is even under SD and the second lowest is
odd. For $S=1$, it is vice-versa. The $\Gamma_5$ states can be viewed
as being of $p$-wave type. 

For $S=0$, the lowest mode vanishes at the same
$x$ as does the gap $\Delta$. So no additional instability
occurs for $S=0$. This provides evidence against a competing 
singlet phase as presumed in Ref.~\cite{koga00}. There is, however,
a salient instability for $S=1$ (Fig.~\ref{fig:s1}) at
$x=0.63$. This comes as a surprise since one expects in antiferromagnets
binding effects to be largest for $S=0$. The singularity at $x=0.630(5)$
is very stable occuring in  all non-defective Dlog-Pad\'e approximants.
 We take the vanishing of a bound 2-triplet state
with $S=1$ as evidence for  strong attraction between triplets which are
neither parallel ($S=2$) nor antiparallel ($S=0$). The attraction
points towards a first order transition into a condensate of triplets
occuring at much lower $x$ than thought previously \cite{mulle00a}.
The angle being neither zero nor $\pi$ corroborates a helical
order if one adopts a classical view  \cite{albre96,mulle00a}.
 In the light of the instability for $S=1$
we interprete the findings by Koga and Kawakami \cite{koga00}
 as indication of the same instability. It is so far not
 excluded that the bulk triplet condensate (the helical phase) is 
 a singlet  or that it can be linked to a singlet  \cite{note4}. 

Next we determine $x$ and $J_1$ for SrCu$_2$(BO$_3$)$_2$. The steep 
decrease of the bound $S=1$ state enables us to fix $x$ very precisely.
Based on ESR \cite{nojir99}, FIR \cite{room00} as well as INS \cite{kagey00a}
we assume $\Delta=2.98$meV and $\omega|_{S=1}=4.7$meV leading to
$x=0.603(3)$ and $J_1=6.16(10)$meV. The 1-triplet dispersion is 
in excellent agreement with experiment 
(cf.~inset in Fig.~\ref{fig:s0} and Ref.~\cite{knett00c}). 
Raman scattering \cite{lemme00a} provides further
strong support because the energy of  the
$\Gamma_3$ singlet matches 30cm$^{-1}$ perfectly.
The $\Gamma_4$ singlet at 25cm$^{-1}$ \cite{note5} is forbidden by symmetry
since  the Raman operator is effectively $\Gamma_3$.

In leading order $t/U$ the Raman operator 
$R=\sum\gamma_{{\bf  i},{\bf j}}{\bf S}_{{\bf i}}{\bf S}_{{\bf j}}$
couples the same spins  as the Hamiltonian. But only the antisymmetric
($\gamma_1=-\gamma_2$) part of $R$ on the dashed bonds (cf.~Fig.~\ref{fig:raman})
creates excitations from the ground state. By geometry we have $\gamma_1''=\gamma_2$
and $\gamma_2''=\gamma_1$  so that the effective $R_{\rm eff}$ is
odd under R1 and R2. 
But the projection
of the vector potential ${\bf A}$ (${\bf A} || {\bf E}$)
 on the bonds under study
in the polarisation ({\bf ab}) implies
$\gamma_1=\gamma_2=-\gamma_1'=-\gamma_2'=\gamma \cos(2\alpha)$\
($\gamma$ microscopic constant, $\alpha$ angle in Fig.~\ref{fig:raman}), i.e.\
an even component implying  $R_{\rm eff}=0$. On the contrary, polarisation 
({\bf a'b'}) yields 
$\gamma_1=-\gamma_2=-\gamma_1'=\gamma_2'=\gamma \sin(2\alpha)$ so that
$R_{\rm eff}\neq0$. This finding agrees nicely with experiment
where on $T\to 0$ the intensities almost vanish for ({\bf ab}) but
grow for ({\bf a'b'}) \cite{lemme00a}. Additionally, $\gamma_1=-\gamma_1'$
and  $\gamma_2=-\gamma_2'$ imply odd parity under SD so that $R_{\rm eff}$
is indeed $\Gamma_3$, not $\Gamma_4$.
Calculating the next $\Gamma_3, S=0$ bound state
(less systematically) yields 45cm$^{-1}$ in good agreement with the
experimental 46cm$^{-1}$ line, too.

We conclude that the 2D model (\ref{hamilton}) explains the low-lying
excitations of SrCu$_2$(BO$_3$)$_2$ perfectly. 
Thermodynamic quantities like the susceptibility
$\chi(T)$ require the inclusion of the interplain coupling $J_\perp$
which is fully frustrated not changing the dimer spins
\cite{miyah00b}. We have employed a Dlog-Pad\'e approximant for the
$1/T$ expansion of the 2D $\chi_{\rm 2D}$
 \cite{weiho99a} complemented
by the condition $\Delta=2.98$meV. This ansatz works fine for $T>35$K.
The 3D $\chi_{\rm 3D}$ is computed
from $\chi_{\rm 2D}$ on the  mean-field level as
$\chi_{\rm 3D}^{-1}=\chi_{\rm 2D}^{-1}+4J_\perp$. The inset in Fig.~\ref{fig:s1}
shows that theory ($J_\perp=1.3(2)$meV) and   experiment \cite{kagey99} 
agree without flaw above 40K.
Our value for $J_\perp$ is significantly higher than the one in Ref.~\cite{miyah00b}
due to different values of $x$ and $J_1$.

The above
comprehensive analysis of bound states is a fine example
of the efficiency of perturbation by unitary transformation  \cite{knett00a} based on
flow equations. This clear concept allows to 
distinguish uniquely sectors with different particle numbers and other
different quantum numbers like the total spin.
Here the concept was put to use to analyse the
Shastry-Sutherland lattice as realized in SrCu$_2$(BO$_3$)$_2$.
 To our knowledge it is the first
 quantitative description of 2-particle bound states in 2D.

The symmetries of experimentally relevant states were determined.
The reliability of the high order results allows to fix the 
experimental coupling constants very precisely 
($J_1=6.16(10)$mev, $J_2/J_1=0.603(3)$, $J_\perp=1.3(2)$meV). 
Thereby, different experiments (ESR, FIR, INS, Raman, $\chi(T)$) 
are explained consistently.
We suggest to exploit the double degeneracy derived here
to resolve different bound states at the border of the MBZ.

An unexpected instability for the $S=1$ 2-triplet bound state
  is found at $x\approx0.63$ indicating a transition to
a triplet condensate probably related to the helical phase found previously
\cite{albre96,mulle00a}. We conjecture that this transition is first
order  occuring at lower $x$ than assumed so far. The vicinity of 
SrCu$_2$(BO$_3$)$_2$ to this transition suggests to attempt a closer
experimental analysis. Pressure and/or substitution will  certainly influence
the ratio $J_2/J_1$. Thereby one may hope to scan through the transition
and to examine the phase beyond.

The authors like to thank H.~Kageyama and P.~Lemmens for generous 
provision of data prior to publication and discussion. The work is supported by
the DFG in SFB 341 and in SP 1073.


\begin{thebibliography}{10}
\bibitem{note5}
The small shoulder at 25cm$^{-1}$ may be a trace of the $\Gamma_4$ singlet
due to higher order processes or a singlet-triplet excitation.
\end{thebibliography}

\end{document}